\documentclass[10pt,english,aps,prl,showpacs,twocolumn]{revtex4}
\usepackage{amsmath}
\usepackage{amssymb}
\usepackage{graphicx}
\usepackage{epsfig}
\usepackage{psfig}

\begin{document}

\title{Activated events in glasses: the structure of entropic droplets}
\author{Maxim Dzero$^1$, J\"{o}rg Schmalian$^1$ and Peter G. Wolynes$^2$}
\affiliation{$^1$Department of Physics and Astronomy and Ames Laboratory, Iowa State
University, Ames, IA 50011}
\affiliation{$^2$Department of Chemistry and Biochemistry and Department of Physics,
University of California, San Diego, La Jolla, CA 92093 }
\date{\today}

\begin{abstract}
Using an effective potential approach, we present a replica instanton
theory for the  dynamics of entropic droplets in glassy systems. 
Replica symmetry breaking in the droplet interface leads to a length
scale dependent reduction of the droplet surface tension and changes the
character of the dynamical heterogeneity and activated dynamics in glasses.
\end{abstract}

\maketitle

At it's most basic level the puzzle presented by the existence of glasses is
\textquotedblleft how can any disordered arrangement of atoms be
stable?\textquotedblright . Disorder implies myriads of structural
possibilities - why don't the unguided thermal motions of the atoms in a
glass take advantage of these possibilities and lead to ceaseless change and
flow? Actually, glasses do flow, but only very slowly. Thus our na\"{\i}ve
question must take on a more quantitative form- \textquotedblleft why do
materials with apparently high entropy density flow as slowly as they
do?\textquotedblright\ Molecules in ordinary glass undergo displacements 
greater than an angstrom only on time scales greater than  hours.
Answering this quantitative question is tantamount to a theory of the glass
transition.

The most successful treatment of this issue to date is provided by the
random first order transition (RFOT) theory of supercooled liquids and
glasses\cite{KW87,KW87b,KTW89,Mon95,MP991,Franz95,Franz97}. Mean field calculations on these systems
suggest a discontinuous
transition without latent heat at a Kauzmann temperature $T_{K}$,  as
found in infinite range spin glasses that lack reflection symmetry\cite%
{Gross85}, which is called ``random first order''. 
The mean field approaches include the use of replicas\cite%
{KW87b,Mon95,MP991,Franz95,Franz97,Gross85,Sommers} and dynamical theories\cite{KT87,CK93} and exhibit
1-step replica symmetry breaking at $T_K$ and a dynamical crossover temperature, $%
T_{A}>T_{K}$, for the onset of activated dynamics. Although a system with infinite
range forces  can break ergodicity while retaining a macroscopic
configurational entropy, this is impossible for a system with finite range
interactions\cite{KW87b,Biroli04}.  Ergodicity will be
restored by \textquotedblleft entropic droplets\textquotedblright , in which
a region of the frozen solid is replaced by any of an exponentially large
number of alternatives. The entropy due to these many alternatives acts as a
driving force for structure change which is, however, off-set by the free energy cost
of matching two alternative structures at their boundaries. This conflict
gives the free energy barrier for slow activated motions.

Many quantitative aspects of the glassy dynamics of molecular liquids 
have been predicted
using the RFOT theory and are found to be in good agreement with
experiment. The predictions include the relation between both the mean
activation barriers in the liquid and glass regimes\cite{XW00} 
and the non-exponentiality of relaxation\cite{XW01} with the
configurational heat capacity, the size scale of the mosaic structures\cite%
{XW00}, which show up as \textquotedblleft dynamical
heterogeneities\textquotedblright\ in the laboratory\cite{exptheter}, as
well as a quantitative prediction of the anomalous low temperature
properties of glasses, including their linear specific heat\cite{Lubchenko01}%
. These  successes of the RFOT theory encourage us to put the
approximations presently used on firmer formal ground. This will allow 
the extension of the RFOT theory to systems more exotic than structural glasses
such as colloidal systems having
stripes or lamellae, i.e.  micro-emulsions, which are predicted to be in the same
universality class\cite{Wu02}, or to  glassy phases with strongly quantum
degrees of freedom such as the electronic motions of correlated electron
systems\cite{SW00}.

In this paper, using replica based methods,
 we study  droplet dynamics in a frozen glassy system
and show how it  differs fundamentally from nucleation in
ordinary first order transitions. These calculations yield a temperature and length scale dependent
reduction of the droplet surface tension that strongly affects the dynamical
heterogeneity and activated dynamics of glassy systems. Specifically, we
show that the transition states in a statistical field theory are instantons
possesing additional replica symmetry breaking (RSB) in the entropic 
droplet and argue that this new RSB state arises from configurational
entropy fluctuations. Instanton like theories that ignore this complexity, particularly  at the droplet interface have been presented earlier\cite%
{KW87b,Parisi94,Takada97,Franz04}. Our results demonstrate that those
approaches contain instabilities  while the present RSB solution strongly supports the wetting picture put forward in\cite{KTW89}.

We consider a system of interacting particles that becomes glassy at low
temperatures. Recent progress in the replica description of glass-forming
systems with entropy crisis\cite{Mon95,MP991,Franz95} allows us to formulate
the RFOT theory in terms of a spatially varying collective variable, $q_{ab}\left( r\right) $,
that corresponds to the local Debye Waller factor of molecular motions for liquid systems.  $q_{ab}=0$ in the
ergodic liquid while $q_{ab}\rightarrow{O(1)}$ for $a\neq b$ in case
of perfect long time correlations. Within replica mean field theory\cite%
{Mon95,MP991,Franz95} the long  time density correlations approach $q\rho
_{0}S\left( k\right) $, with the equilibrium liquid structure factor being $S\left( k\right) $,
the mean particle density $\rho _{0}$ and the  off-diagonal element $q$ of $q_{ab}$.
Below the dynamic transition temperature $T_{A}$, $q$ becomes nonzero for the first time
and an extensive configurational entropy $S_{c}$ occurs, that vanishes
linearly at the Kauzmann temperature $T_{K}<T_{A}$. 
A heterogeneity of the glass, i.e. the emergence of correlated long 
time dynamics over certain length scales, is reflected in an $r$-dependence of $q_{ab}(r)$.

The mean field value and fluctuations of $q_{ab}$ are determined by an
effective Hamiltonian, $H\left[ q\right] $. In order to keep our calculation
transparent we will not specifically determine $H\left[ q\right] $ 
from an explicit calculation for supercooled liquids (for an example see Ref.\cite{Franz97}), but
start from a simple polynomial  model in the same universality class\cite{Gross85}: 
\begin{equation}
H=\sum_{a,b}\int d^{d}r\left( h\left[ q_{ab}\right] -\frac{1}{3}%
\sum_{c}q_{ab}q_{bc}q_{ca}\ \right)   \label{HH}
\end{equation}%
with $h\left[ q_{ab}\right] =\frac{1}{2}\left( \nabla q_{ab}\right) ^{2}+%
\frac{t}{2}q_{ab}^{2}-\frac{1+w}{3}q_{ab}^{3}+\frac{y}{4}q_{ab}^{4}$. In Eq.%
\ref{HH} length scales are measured in units of the interparticle distance
and energy scales in units of a scale that determines the absolute value of
the configurational entropy. The parameters $t$, $w>0$ and $y>0$,
which are in principle all temperature dependent, become dimensionless. We
take the primary $T$-dependence to reside in the quadratic term, $t=%
\frac{T-T_{0}}{T_{0}}$ where $T_{0}$ determines the temperature scale where
slow dynamics of the model sets in. We use both the absolute and reduced
temperature variables, $T$ and $t$, i.e. $T_{K}$ or $t_{K}=t\left(
T_{K}\right) $ etc.

Our aim is to perform an instanton calculation to describe the escape from a given 
metastable frozen solid configuration. This requires a technique that agrees in the
homogeneous limit with other mean field calculations within the RFOT
universality class\cite{MP991,SW00}, but allows us to study the behavior for
arbitrary values of the overlap between two states. This is achieved by the
quenched effective potential approach of Franz and Parisi\cite{Franz95}. One considers
a fixed overlap $p_{c}\left( r\right)=\rho_1(r) \rho_2(r) $ between two particle
density configurations and  determines the constrained partition function 
$Z(p_c,\rho_2)=\langle \delta( p_{c}\left( r\right)-\rho_1(r)\rho_2(r)) \rangle_{\rho_1} $
by thermally averaging
over one configuration. The effective potential, $\Omega[p_c]=-T \langle \log Z(p_c,\rho_2)  \rangle_{\rho_2} $,
 is the average of the free energy over the second configuration\cite{Franz95} and can be written as:
\begin{equation}
\Omega \left[ p_{c}\right] =2\int d^{d}rh\left[ p_{c}\right] +\ F\left[ p_{c}
\right]. 
\end{equation}
Here $F\left[ p_{c}\right] =-T\left. \frac{\partial }{\partial m}\int
Dq\exp \left( -\beta H_{p_{c}}\left[ q\right] \right) \right\vert
_{m\rightarrow 0}$
is the quenched free energy average for a system with Hamiltonian, Eq.\ref%
{HH}, in an external field $p_{c}^{2}\left( r\right) $: $H_{p_{c}}\left[ q%
\right] =H\left[ q\right] -\sum_{ab}\int d^{d}rp_{c}^{2}\left( r\right)
q_{ab}\left( r\right) $.
\begin{figure}[h]
\vspace{-0.5cm}
\centerline{\psfig{file=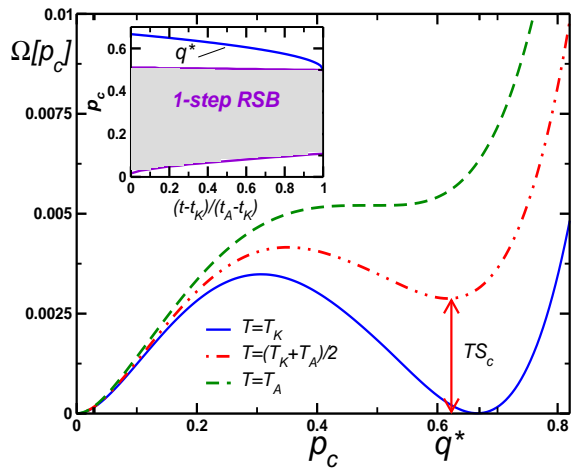,height=8.5cm,width=7cm,angle=-90}}
\caption{Effective potential as a function of the overlap, $p_c$
for different temperature; inset: region of $p_c$ values 
where RSB occurs together with $q^{*}(t)$.}
\label{Fig1}
\vspace{-0.5cm}
\end{figure}
Replica theory gives the statistical interpretation of $p_c$ and $q_{ab}$. 
$p_c$ refers to the similarity with the initial state while $q_{ab}$ measures the configurational
similarity between different states, regardless of the initial state. 

Before we perform the instanton calculation we summarize the homogeneous
mean field theory of Eq.\ref{HH}. In Fig.1 we show $\Omega \left[ p_{c}%
\right] $ for $r$-independent $p_{c}$ where $q_{ab}$ in $F\left[ p_{c}\right]
$ is determined via saddle point method. The minima of $\Omega \left[ p_{c}%
\right] $ determine the mean field value of the Debye Waller factor 
$q^{\ast }=\frac{w+\sqrt{w^{2}-4t}}{2y}$ and the configurational
entropy $\Omega \left[ q^{\ast }\right] =TS_{c}$\cite{Franz95}.
We find $t_{A}=\frac{w^{2}}{4y}$ and $t_{K}=\frac{2w^2}{9y}$ for the
reduced dynamic crossover and Kauzmann temperatures, respectively. The
configurational entropy density, $s_{c}=S_{c}/V$,
vanishes linearly at the Kauzmann temperature, $s_{c}\left( t\simeq
t_{K}\right) =\frac{1}{4T_{K}}q^{\ast 2}\left( t_{K}\right) \left(
t-t_{K}\right) $, while $s_{c}\left( t_{A}\right) =\frac{1}{12T_{A}}q^{\ast
2}\left( t_{A}\right) t_{A}$. \ For $p_{c}=q^{\ast }$, $q_{ab}$ is replica
symmetric. The lowest replicon eigenvalue, $\lambda $, is positive for $%
t_{K}\leq t<t_{A}$ and vanishes at $t_{A}$ like 
$\lambda\propto\left(t_{A}-t\right)^{1/2}$. Since replica symmetry in the effective 
potential formalism corresponds to one step RSB in the
 conventional replica approach\cite{Parisi04}, all these results are in agreement with
the mean field theory of a RFOT. 
For $p_{c}$ values between $0$ and
 $q^{\ast }$ an additional replica symmetry breaking in $q_{ab}$ occurs\cite%
{Sommers,Barrat97} (see inset of Fig.1).

The $\Omega \left( p_{c}\right) $ of Fig.1 clarifies the fact that between $T_{K}$ and $T_{A}$ the frozen
solid with $p_{c}=q^{\ast }$ is metastable with respect to the liquid ($%
p_{c}=0$) and higher in free energy by $TS_{c}$. The system will try
to reduce its energy by realizing this entropy via droplet nucleation. We
argue that the transition state for this nucleation is determined by an
instanton solution for the overlap profile $p_{c}\left( r\right) $,
i.e. via $\frac{\delta \Omega \left[p_{c}\right] }{\delta p_{c}\left(
r\right) }=0$ with barrier $\Omega\left(p_{c}\right)-
\Omega \left( q^{\ast }\right) $ ; see also Ref.\cite{Franz04}

Assuming for the moment replica symmetry for the instanton, $q_{ab}\left(
r\right) =\left( 1-\delta _{ab}\right) q\left( r\right) $ we find the
solution $p_{c}\left( r\right) =q\left( r\right) $, determined by $\Delta
p_{c}\left( r\right) =\frac{dV\left( p_{c}\right) }{dp_{c}\left( r\right) }$
with $V\left( p_{c}\right) =\frac{t}{2}p_{c}^{2}-\frac{w}{3}p_{c}^{3}+\frac{y%
}{4}p_{c}^{4}$. While $V(p_c)$ and $\Omega \left( p_{c}\right) $ differ in detail, they have rather similar shape and agree for $p_c=q^\ast$.
If replica symmetry holds, the problem then becomes almost identical to
ordinary homogeneous nucleation\cite{Langer67} with transition temperature $%
T_{K}$ and spinodal $T_{A}$.  Close to $T_{K}$ a thin interface and a
diverging droplet size result. The competition between surface and bulk
terms determines the barrier 
\begin{equation}
\Delta F=4\pi \sigma _{0}R^{2}-\frac{4\pi }{3}Ts_{c}R^{3}.  
\label{barriesRS}
\end{equation}%
The sole distinction from ordinary nucleation is that the driving force for
the decay is entropic. For the surface tension, the model Eq.\ref{HH} yields
 $\sigma _{0}\simeq
q^{\ast 2}\left( t_{K}\right) t_{K}^{1/2}$, the interface width is of order $%
l_{0} =2^{7/2} t_{K}^{-1/2}$ and the droplet radius is $R\simeq l_{0}\frac{%
t_{K}}{t-t_{K}}$. The time scale for activated dynamics therefore diverges 
like $\log \frac{\tau }{\tau _{0}}\propto \left( \frac{t_{K}}{t-t_{K}}\right) ^{d-1}$
at $T_{K}$ for spatial dimension $d>1$. This result was first found in
Ref.\cite{KW87b} and also given in \cite{Parisi94,Franz04}. Close to 
$T_{A}$ no sharp droplet interface can be defined\cite{Unger84}. A
dimensional analysis of the instanton equation near $T_A$ yields the
length $\xi _{A}\simeq \left( t_{A}\left( t_{A}-t\right) \right)
^{-1/4}$, identical to the length scale
in mode coupling theory above the upper critical dimension 
$d_{uc}=6$\cite{Biroli04b}, see also Ref.\cite{KW87b}. 
Recently, using the effective potential approach, all these results for replica
symmetric $q_{ab}$ were independently obtained by Franz\cite{Franz04}.

The self-generated randomness of a glassy system however should modify the nucleation dynamics.
A crucial question is whether the replica symmetry
for $q_{ab}$ is indeed stable for the droplet. A stability analysis shows that
the replicon eigenvalues, $\lambda $, of the droplet solution
are determined by 
\begin{equation}
\left( -\nabla ^{2}+U\left( r\right) \right) \psi \left( r\right) =\lambda
\psi \left( r\right) ,
\end{equation}%
with $r$-dependent potential $U\left( r\right) =\left. \frac{d^{2}V\left(
q\right) }{dq^{2}}\right\vert _{q=q^{\ast }}$.  Without the gradient term
we would recover the result for $\lambda $ discussed above and replica symmetry is
stable below $T_{A}$. The stability changes if we include, however, the gradient term.
Plainly, the replicon eigenvalues are identical to the eigenvalues of the
usual fluctuation determinant in the decay of a metastable state.
Thus, there is one negative eigenvalue for 
$T_{K}<T<T_{A}$. Close to $T_{K}$ it follows that $\lambda\simeq{-R^{-2}}$ with
droplet radius $R$. The unstable eigenvector is in the replicon direction and this
instability occurs \emph{in addition} to the expected instability with
respect to the growth mode, i.e. the motion of the interface of the droplet.
We further infer that the replicon
eigenfunction, $\psi \left( r\right) $, is localized on the droplet
interface.

The instability of the replica symmetric solution entails an additional
replica symmetry breaking as the overlap $q_{c}$ goes "over the hill", i.e.
close to the interface. Allowing more general solutions for $q_{ab}\left(
r\right) \rightarrow q\left( r,x\right) $, where $x\in \left[ 0,1\right] $
is the usual RSB parameter, we obtain from $\frac{\delta \Omega \left[ p_{c}%
\right] }{\delta p_{c}\left( r\right) }=0$ and the saddle point equation
for $q_{ab}$, two coupled instanton equations: 
\begin{equation}
\nabla ^{2}p_{c}\left( r\right) =\frac{dV\left( p_{c}\right) }{dp_{c}\left(
x\right) }\ -p_{c}^{2}\left( r\right) \ +p_{c}\left( r\right)
\int_{0}^{1}q\left( r,x\right) dx
\label{eq6}
\end{equation}
and 
\begin{equation}
\begin{split}
\nabla ^{2}q\left( r,x\right) &=\frac{dV\left( q\right) }{dq\left(
r,x\right) }-\left( 1-x\right) q^{2}\left( r,x\right) -p_{c}^{2}\left(
r\right)  \\
&+\int_{0}^{x}dx^{\prime }q^{2}\left( r,x^{\prime }\right) +2q\left(
r,x\right) \int_{x}^{1}q\left( r,x^{\prime }\right).
\end{split}
\label{eq7}
\end{equation}
We have numerically solved these equations under the assumption of spherical
symmetry and with the approximation that $\nabla ^{2}\simeq \frac{d^{2}}{%
dr^{2}}$, justified for large droplet radius, i.e. for $T$ close to $T_K$.
 We find that new solutions
with one step replica symmetry breaking exist! In Fig.2 we show our results
for $p_{c}\left( r\right) $ and for $q\left( r,x\right) $ below ($%
q_{0}\left( r\right) $) and above ( $q_{1}\left( r\right) $) the breakpoint, 
$\overline{x}$.  We find $\overline{x}$ to be $r$- independent. Outside the
droplet, $p_{c}$ and the induced density correlations $q\left( r,x\right) 
$ all approach the mean field value $q^{\ast }$.  However, inside
of it, $q(r,x)$ has a nontrivial distribution function. With weight
$\overline{x}$, $q(r,x)$ is small ($q_{0}\lesssim p_{c}$) and the transition state structures are
configurationally distinct.  However, with smaller weight, $1-\overline{x}$, a few energetically well matching states
contribute to making the transition.
 This demonstrates that dynamical heterogeneity in glasses differs
fundamentally from ordinary nucleation. 
This instability originates in the interface, where the gradients are nonzero, but affects, via boundary conditions,
 the overlap distribution in the entire droplet.
As $T\rightarrow T_{K}$ we find $\overline{x}\rightarrow 1$.
At the Kauzmann temperature, where the replicon eigenvalue vanishes, the system is marginal in a new, heterogeneous sense.
\begin{figure}[h]
\vspace{-0.4cm}
\centerline{\psfig{file=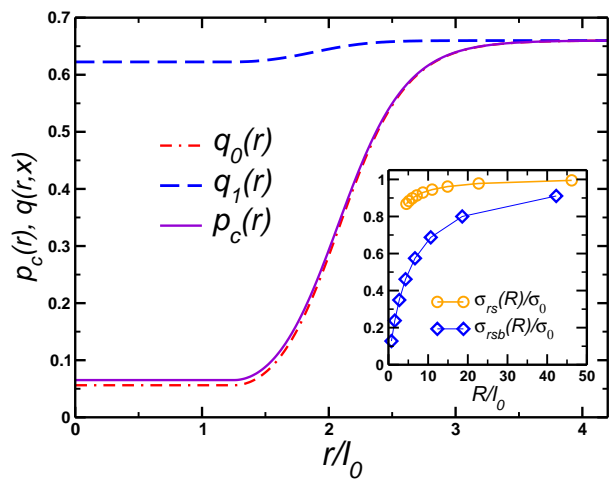,height=9cm,width=7cm,angle=-90}}
\caption{Instanton profile for $t=t_K+0.08(t_A-t_K)$ and $y=w=1$ with breakpoint $\overline{x}=0.975$ and 
$q_{0,1}=q(x \lessgtr \overline{x},r)$. Inset: surface tension as function of the droplet radius 
for the RS and RSB instantons.}
\label{Fig2}
\vspace{-0.3cm}
\end{figure}

If we insert our solution into $\Omega \left[ p_{c}\right] $ we
obtain the interesting result that, compared to the replica symmetric case,
the bulk driving force for nucleation is unchanged. The new solution does
however change the value of the surface tension. Close to $T_K$ our numerical results are
very well described by $\sigma \left( t\right) =\sigma _{0}\left( 1-b\frac{%
t-t_{K}}{t_{A}-t_{K}}\right) $ with large values for the constant $b$ ($%
b\simeq 10$ for $w=y=1$ and $b\simeq 5$ for $w=3$, $y=1$). The reduction of the
surface tension, shown in the inset of Fig. 2 as function of droplet radius 
$R=\frac{2\sigma \left( t\right) }{Ts_{c}\left( t\right) }$, is
caused by the fact that RSB recognizes the existence of structures whose interface matches
are less costly.

The driving force for the decay of the metastable frozen solid is entropic. 
As usual for a replica based theory, $S_c$ is only the mean value of
the logarithm of the number of states.  While fluctuations of
the entropy density vanish for infinite systems, $\overline{\left( \delta
s_{c}\right) ^{2}}=c_{c}/V$, they become relevant in finite subsystems like
small droplets. For example we
find for the configurational specific heat capacity  at $T_{K}$:  $c_{c}=
\frac{2q^{\ast 3}\left( t_{K}\right) }{3T_{K}}$. Configurational entropy fluctuations, i.e. fluctuations of the
driving force for nucleation,
act similarly as do  field fluctuations in the random field Ising model (RFIM). In 
the RFIM simple statistical arguments yield a reduction of
the surface tension
 $\sigma \left( R\right) =$ $\sigma _{0}\left( 1-const. \, R^{-3/2}  \right) $. 
 The effect is caused by fluctuations, 
corresponding to a wetting of the interface by states that 
make the matching between inside and outside the drop easier.
 $\sigma \left( R\right)$ computed this way behaves very similar to our numerical results presented above.
In the context of the RFIM, Villain\cite{Villain} proposed
that one can consider such a $\sigma \left( R\right) $ as the first step of
a renormalization group argument to the surface tension and obtained  $\sigma \propto R^{-\frac{d-2}{2}}$ from the integration of the flow
equation. For an RFOT this leads to a droplet radius $R^{\ast }\propto \left(
t-t_{K}\right) ^{-2/d}$ and \ a Vogel-Fulcher law: 
$\tau =\tau _{0}\exp \left[DT_{K}/(T-T_{K})\right]$ 
for the activated dynamics as advocated in Ref\cite{KTW89}. Our results strongly
support this wetting picture in entropic droplets. For practical applications in liquids, 
where it is hard to approach $T_K$ very closely,
the present first step of the full renormalization group calculation, may well be  sufficient.

In summary we have shown that the transition states responsible for the slow
activated dynamics in glassy systems are given by instantons with new replica
symmetry breaking (RSB). This new RSB is due to a changed distribution 
of states at an interface between a frozen solid and a state which 
explores various metastable states. 
We note, the present treatment also allows 
naturally the inclusion of effects that will lead to barrier height fluctuations and
non-exponential glassy relaxation.

We thank Jean-Philippe Bouchaud for helpful discussions. This research was
supported by the Ames Laboratory, operated for the U.S. Department of Energy
by Iowa State University under Contract No. W-7405-Eng-82 (M.D. and J. S.),
a Fellowship of the Institute for Complex Adaptive Matter (M.D.), and the
National Science Foundation grant CHE-0317017 (P. G. W.).


\bigskip

\begin{thebibliography}{99}
\bibitem{KW87} T. R. Kirkpatrick and P. G. Wolynes, Phys. Rev. A \textbf{35}%
, 3072 (1987).

\bibitem{KW87b} T. R. Kirkpatrick and P. G. Wolynes, Phys. Rev. B \textbf{36}%
, 8552 (1987).

\bibitem{KTW89} T. R. Kirkpatrick and D. Thirumalai, and P. G. Wolynes,
Phys. Rev. A \textbf{40}, 1045 (1989).


\bibitem{Mon95} R. Monasson, Phys. Rev. Lett. \textbf{75}, 2847 (1995).

\bibitem{MP991} M. Mezard and G. Parisi, Phys. Rev. Lett. \textbf{82}, 747
(1999).

\bibitem{Franz95} S. Franz and G. Parisi, J. Phys. I (France) \textbf{5},
1401 (1995).

\bibitem{Franz97} S. Franz and G. Parisi, Phys. Rev. Lett. \textbf{79}, 2486
(1997).


\bibitem{Gross85} D. J. Gross, I. Kanter, and H. Sompolinsky, Phys. Rev.
Lett. \textbf{55}, 304 (1985).

\bibitem{Sommers} A. Crisanti and H.-J. Sommers, Z. Phys. B: Cond. Mat. 
\textbf{87}, 341 (1992).

\bibitem{KT87} T. R. Kirkpatrick and D. Thirumalai, Phys. Rev. Lett. \textbf{%
58}, 2091 (1987).

\bibitem{CK93} L. F. Cugliandolo, J. Kurchan, Phys. Rev. Lett. \textbf{71},
173 (1993).

\bibitem{Biroli04} G. Biroli and J.-P. Bouchaud, Journ. of Chem. Phys., 
\textbf{121}, 7347 (2004).

\bibitem{XW00} X. Xia and P. G. Wolynes, Proc. Natl. Acad. Sci. \textbf{97},
2990 (2000);V. Lubchenko, P. G. Wolynes, \ Journ. of Chem. Phys. \textbf{121}%
, 2852 (2004).


\bibitem{XW01} X. Xia and P. G. Wolynes, Phys. Rev. Lett. \textbf{86}, 5526
(2001).

\bibitem{exptheter} U. Tracht \emph{et. al.}, Phys. Rev. Lett. \textbf{81}, 
2727 (1998); E. V. Russel and N. E. Israeloff, Nature {\bf 408}, 695 (2000);
X. H. Qiu and M. D. Ediger, Jour. Phys. Chem. B {\bf 107}, 459 (2003).


\bibitem{Lubchenko01} V. Lubchenko, P. G. Wolynes, Phys. Rev. Lett. \textbf{%
87}, 195901 (2001).

\bibitem{Wu02} S.Wu, H. Westfahl Jr., J. Schmalian, and P. G. Wolynes, Chem.
Phys. Lett. \textbf{\ 359}, 1 (2002).

\bibitem{SW00} J. Schmalian and P. G. Wolynes, Phys. Rev. Lett. \textbf{85},
836 (2000);H. Westfahl Jr., J. Schmalian, and P. G. Wolynes, Phys. Rev. B 
\textbf{64}, 174203 (2001).

\bibitem{Parisi94} G. Parisi, preprint, cond-mat/9411115.

\bibitem{Takada97} S. Takada, P. G. Wolynes, J. Chem. Phys. \textbf{107},
9585 (1997).

\bibitem{Franz04} S. Franz, preprint cond-mat/0412383. Of the two approaches
discussed, the quenched one yields a description for entropic
droplets consistent with RFOT. 

\bibitem{Parisi04}G. Parisi, G. Ruocco, and F. Zamponi, Phys. Rev. E {\bf 69}, 061505 (2004).

\bibitem{Barrat97} A. Barrat, S. Franz, and G. Parisi, J. Phys. A: Math.
Gen. \textbf{30}, 5593 (1997).

\bibitem{Langer67} J. S. Langer, Annals of Physics \textbf{41}, 108 (1967).

\bibitem{Unger84} C. Unger and W. Klein, Phys. Rev. B \textbf{28}, 445
(1983).

\bibitem{Biroli04b} G. Biroli and J.-P. Bouchaud, Europhys. Lett. \textbf{67}%
, 21 (2004). 

\bibitem{Villain} J. Villain, J. Physique \textbf{46}, 1843 (1985).
\end{thebibliography}
\end{document}